\documentstyle[12pt,epsf]{article}
\begin{document}
\begin{center}
{\bf TRANSITION TO THE REGION OF CENTRAL COLLISIONS}
\vskip 5mm
M.K. Suleymanov${^1}{^\star}{^\circ} $, O.B. Abdinov${^1} {^\star}$,N.S. Angelov${^2}$, A.I. Anoshin${^3}$,A.S. Vodopianov${^2}$,A.A. Kuznetsov${^2}$
and Z.Ya. Sadigov${^1}{^\star}$
\vskip 5mm
{\small
(1) {\it
Institute of Physics,Academy of Sciences of Azerbaijan Republic}\\
(2) {\it
Laboratory of High Energies,
Joint Institute for Nuclear Research,
141980, Dubna, Moscow Region, Russia}\\
(3) {\it
Nuclear Physics Institute of Moscow State Univerisity,Russia}\\
$\star$ {\it
Now : Laboratory of High Energies,
Joint Institute for Nuclear Research,
141980, Dubna, Moscow Region, Russia}\\
$\circ$ E-mail: mais@sunhe.jinr.ru
}
\end{center}
\vskip 5mm
\begin{center}
\begin{minipage}{150mm}
\centerline{\bf Abstract}

The  experimental results on the behaviour of the characteristics of
secondary particles  depending on the disintegration degree of nuclei
are used to determine the region of central collisions. It was
therefore possible that :

  -- the correlation between the prosesses of total disintegration of
nuclei and the central collisions of nuclei had been shown;

  -- the existence of the regime change points in the behaviour of
the considered characteristics of secondary particles depending on the
disintegration degree of nuclei had been observed in the other
earlier experiments as well.

The number of all protons  in  $ {} ^ {12} CC $- interactions
at the momentum of 4.2 A GeV/c
obtained from the 2-m propane bubble chamber exposed at the Dubna
machine is considered as a disintegration degree of nuclei.

The experimental results demonstrate that there are  cases
corresponding  to the critical  phenomena among the events
with  the  central collisions of nuclei. For $^{12}CC$-interaction
the behaviour of  the number of the events, depending on $Q$ also depends on
the number of fragments and has a two--steps form. This result  could be explained  by the
existence of nuclear clusters.

\end{minipage}
\end{center}
\vskip 10mm
\section{Introduction}

The talk is dedicated to the experimental selection of  the events with total
disintegration of nuclei  or with central collisions of ones. In paper~\cite{1} it was experimentally shown that the
processes with total disintegration of nuclei correspond to the
central collisions. Therefore it can  generalize the rezults of the
experiments on the total disintegration of nuclei and  the central
collisions.  First of all  it is necessary to remind that at present there are
many papers in which the processes of nuclear fragmentation~\cite{2}
and the processes of total disintegration of  nuclei~\cite{3}
are considered as a critical phenomenon.  Therefore it is
possible to  suppose that  if  there are  cases corresponding
to the critical phenomena among the events with  the total
disintegration of nuclei then the points of the regime changes in
the behaviour  of  some characteristics of secondary
particles ($a_i$) depending on the centrality degree of
collisions - $Q$ could be observed.  Some experimental data
obtained in nuclear-nuclear interactions at high energy
demonstrate the existence  of the points  of the regime  changes
in the behaviour of the $a_i =f(Q)$ distributions.  For example
in fig. 1 the averege multiplicity of relativistic chaged
particles depending  on $Q$ is shown for the  $^{28}Si_{14} + Em$
reactions at the energies 3.7  and 14.6 GeV per nucleon. To
determine the $Q$ a number of charged projectile fragments
($Z_{f_i}$)  were used. The figures were obtained  from paper~\cite{4}.
The points of the regime change are observed  in these
dependences. These points were used by the authors to select the
events with central collisions of nuclei.

  In fig. 2 are
shown the averege values of  pseudorapidity - $\eta (\eta = -ln~tg(\theta /2))$
for s-particles(the particles with  $\beta > 0.7$)  depending on the
number of $g$-particles( the  particles with $\beta \leq 0.7$ ) for
$pEm$-reactions at  the momenta of $p_0=$4.5; 24.0;50.0;67.0 and
200.0 GeV/c. This figure was obtained from  paper~\cite{5}. The
dashed line in the figure corresponds to the
cascade-evaporation model calculations. The points of the regime
changes in these distributions are also seen. The
cascade-evaporation model  calculations cannot describe
numerically  these distributions.

In fig. 3 the
$Q$-dependences  of the absolute values of  one-particle correlation
function $\mid R(pt,Y)\mid$ are shown for the protons in the
$^{12}CC$-interactions at the momentum of 4.2 A GeV/c. The data was
obtained from paper~\cite{6}.  Here $Q$ is a number of protons in an
event.  The points of the regime changes in these figures are also seen.
Thus the results clearly demonstrate that  there exist  the points of
the regime change of the behavior of $a_i = f(Q)$ distribution which
could fix the  region of the  central collisions of nuclei.  We believe
that these points could be used to select the events with the central
collisions of nuclei.  So to find the points of  the regime changes it
is possible to investigate the $a_i= f(Q)$  distributions'
behaviour.  The $Q$  could be determinated as:

 - a number  of fragments($n_f$);

 - a number of protons in an event($n_p$);

 - an energy flow  of particles at the  emission  angulars
 $\theta \simeq 0(E_{ZDC} )$ or  $\theta \simeq 90 (E_t)$  ones.

  	For further  confirmation of the results on the existence the
regime change  points and  for clarification of the reasons  of these
points appearance we studied the influence of  nuclear fragmentation
process on the behaviour of the events number depending on $Q$.
Supposing these points beeing connected with  the appearence  of a
critical phenomenon the fragment number change could also
have a critical character with the $Q$ increase. Because the
intermediate nuclear formations (for example nuclear cluster) could be a
source of the nuclear fragmentation.

\vskip 10mm

\section{Experiment}

We used  20407 events of  $^{12}CC$- reactions  at the momentum of 4.2
A GeV/c obtained from the 2-m propane bubble chamber of  LHE, JINR(the
methodical details are described in  papers~\cite{7}).To determine
the values of  the $Q$ two variants were considered.  In the first
variant the values  of the $Q$ were determined as

\begin{equation}
Q=n_+ + N_p -n_- ,
\end{equation}
here $n_+, n_-$  and  $N_p$ are the numbers of identified $\pi^+, \pi^-$-mesons
and protons respectively. In that determination the $Q$ is a number of all
the protons in an event without taking into account a remainder  of nuclei.
In the second variant the Q values were determined as

\begin{equation}
Q=N_+ -n_-      ,
\end{equation}
here $N_+$  are  charges  of
all the positively charged particales in an event including the
nuclear fragments. In that determination the $Q$ is a summary charge of
an event.

\section{Results}

The distributions of the events number depending on the $Q$
are shown in fig.4a,b. In fig.4a  the empty starlets
correspond to the first variant and the full starlets - to
the  second variant.  From fig. 4a  it is seen that  the
fragments number  to determine the $Q$ being included the form
of distributions sharply changes and  has a two-steps
structure.

In fig.4b are shown the $Q$-dependences of the
events  number for the calculation data obtained from the
quark-gluon string model~\cite{8} (QGSM) without the nuclear
fragments. The empty starlets correspond to the cases in
which the stripping protons were not taken into account and
the full starlets correspond to the cases in which the
stripping protons were included. It is seen that the form of
the distribution strongly differs from the experimetal one in
fig.4a. There is no  two-steps structure in this figure.
Therefore we can assert that this difference is connected
with the existence of fragments in experimental events.\\
Thus, the  results demonstrate that the influence of  nuclear
fragmentation process in the behaviour of  the events number
depending on the $Q$ has a critical character. To explain this
result we suppose that it could be connect with the existence
of nuclear clusters.It is possible to think that with the
increasing of centrality degree the probability of claster
formation grows but further increasing the Q ( in the region of
high Q) leads to  the big clasters decay  on nuclear
fragments and then on  free nucleons(see fig.5). It could be a reason
for observation of  two-step structure in the distributions.
The first step connected with the formation of  a cluster and
second one with its decay.

\section{Conclusion}

1. The experimental results obtained in the  nuclear-nuclear
collisions at high energy  clearly demonstrate the points of
the regime change of the behaviour of some characteristics of
econdary particles depending on the centrality degree of
collisions. It could  mean that there are  cases
corresponding  to the critical  phenomena among the events
with  the  central collisions of nuclei.

2. For $^{12}CC$-interaction  the behaviour of  the number of events,
depending on $Q$ also depends on the number of fragments and
has a two--steps form which  is not  reproduced by the
calculated data in the framework of the QGS model without
nuclear fragments. This result  could be explained  by the
existence of nuclear clusters.

\newpage
{\Large \bf~~~~~~~~~~~~~~~~~~Captions of figures}
\vspace{10mm}

1. Q-dependences of multiplicity of relativistic
charged particles in Si+Em reactions at the energies of
a) 3.7 GeV per nucleon; b) 14.6 GeV per nucleon.

2. $n_g$-dependences of the averege values of
 $\eta = -ln~tg(\theta /2))$ for $n_s$-particles.

3. Q-dependences of the absolute values of one particle correlation function
for the protons in $^{12}CC$-interaction.

4. Q-dependences of the event number of ${}^{12}CC$-interactions: a) experiment; b)
 quark-gluon string model.

5. Fig.5

\newpage
\begin{minipage}{4cm}

\end{minipage}

\vskip 4cm
\hspace*{-0cm}
\begin{center}
\hspace*{-0.cm}
\parbox{16cm}{\epsfxsize=16.cm \epsfysize=16.cm\epsfbox[5 5 500 500]
{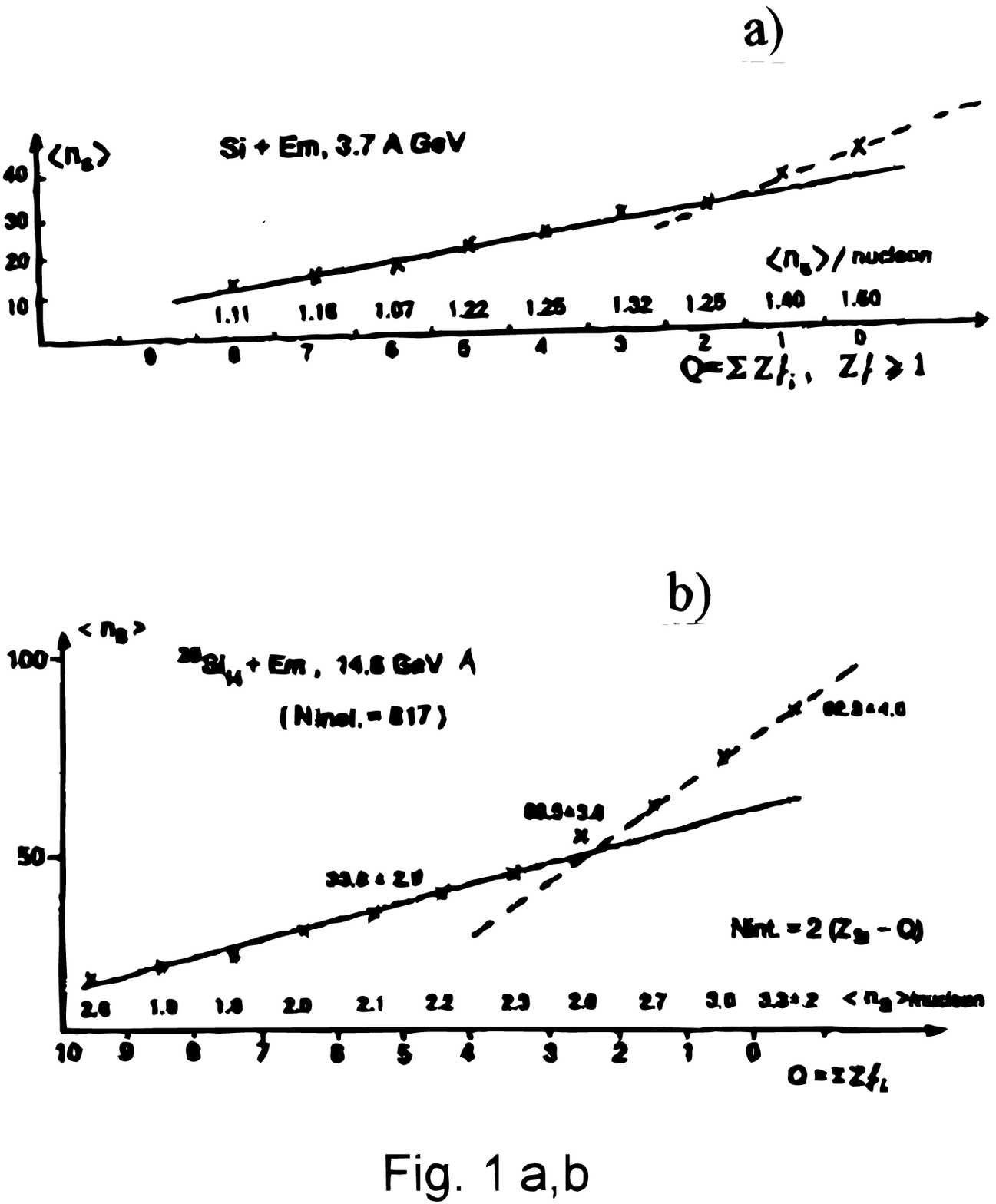}{}}

\vskip 0.25cm

\end{center}


\begin{minipage}{4cm}

\end{minipage}

\vskip 4cm
\hspace*{-0cm}
\begin{center}
\hspace*{-0.cm}
\parbox{16cm}{\epsfxsize=16.cm \epsfysize=16.cm\epsfbox[5 5 500 500]
{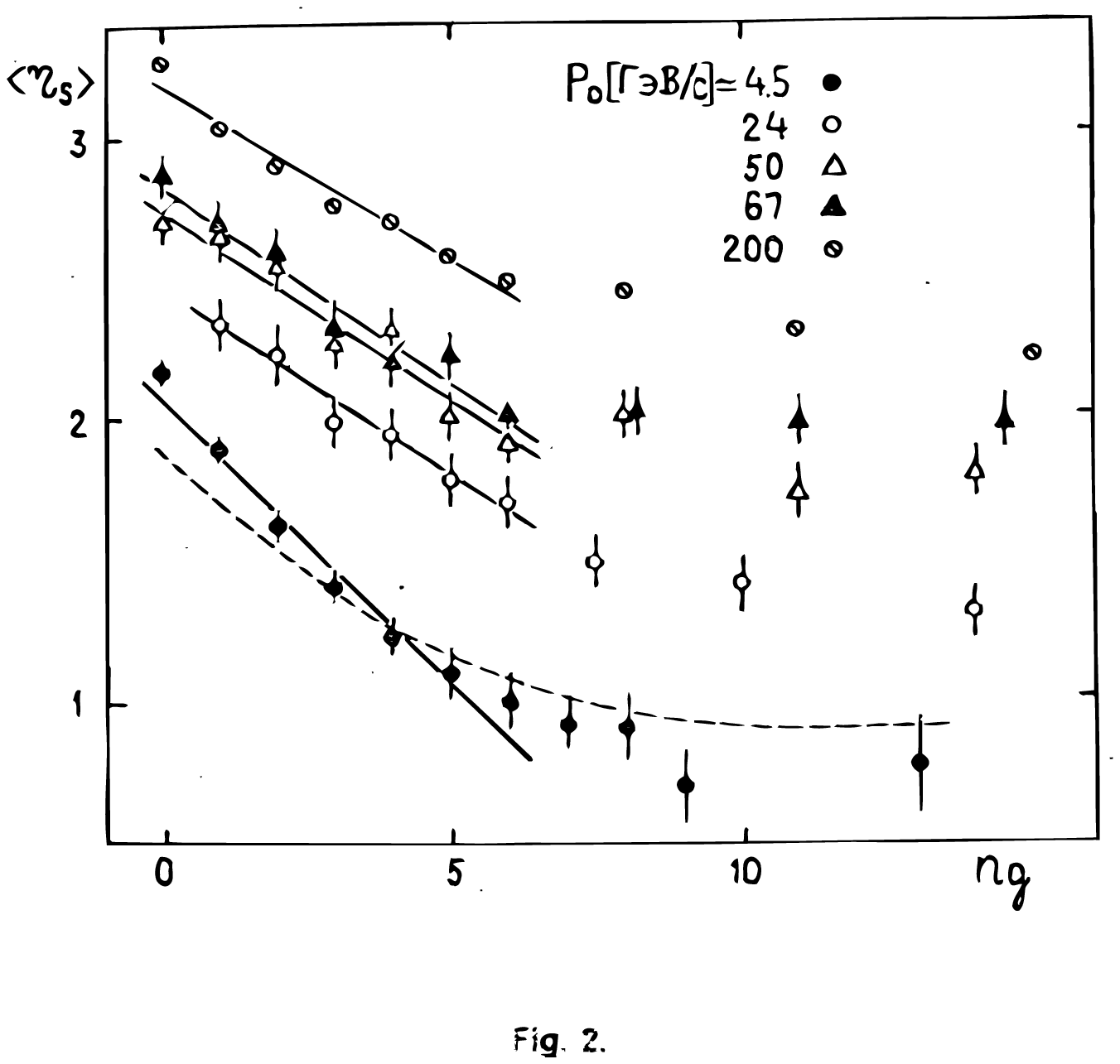}{}}

\vskip 0.25cm

\end{center}
\begin{minipage}{4cm}

\end{minipage}

\vskip 4cm
\hspace*{-0cm}
\begin{center}
\hspace*{-0.cm}
\parbox{16cm}{\epsfxsize=16.cm \epsfysize=16.cm\epsfbox[5 5 500 500]
{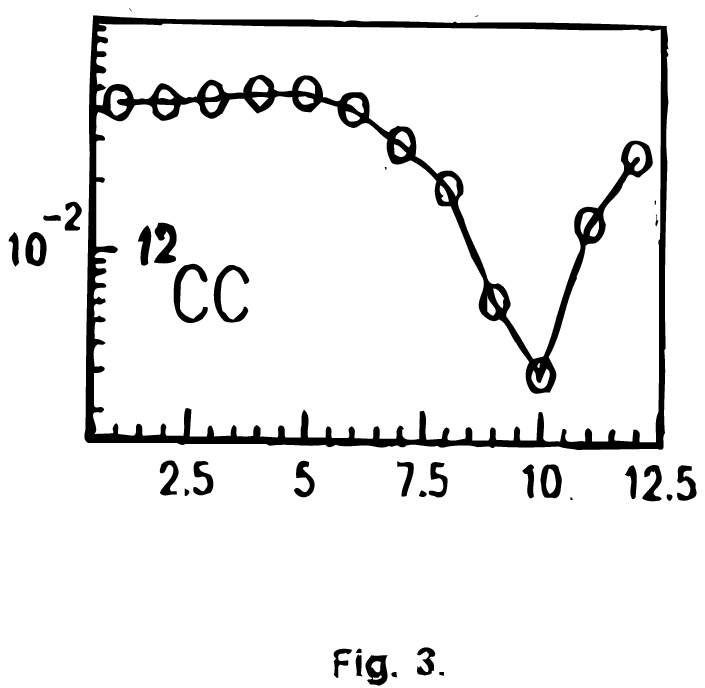}{}}

\vskip 0.25cm

\end{center}
\newpage
\begin{minipage}{4cm}

\end{minipage}

\vskip 2cm
\hspace*{-0cm}
\begin{center}
\hspace*{-0.cm}
\parbox{16cm}{\epsfxsize=16.cm \epsfysize=16.cm\epsfbox[5 5 500 500]
{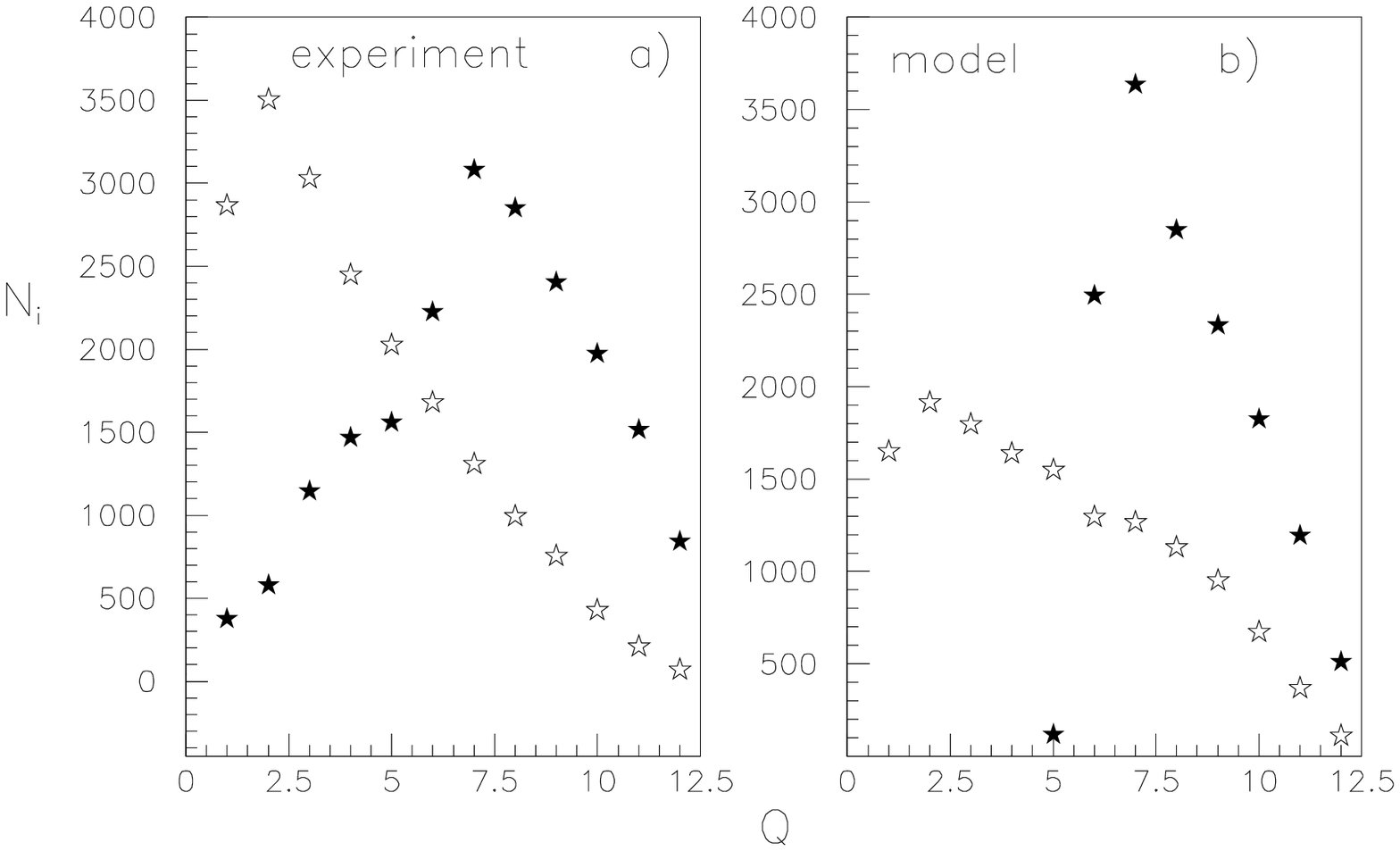}{}}

\vskip 0.25cm

\end{center}
{\bf Figure 4a,b.}

\newpage
\begin{minipage}{4cm}

\end{minipage}

\vskip 4cm
\hspace*{-0cm}
\begin{center}
\hspace*{-0.cm}
\parbox{16cm}{\epsfxsize=16.cm \epsfysize=16.cm\epsfbox[5 5 500 500]
{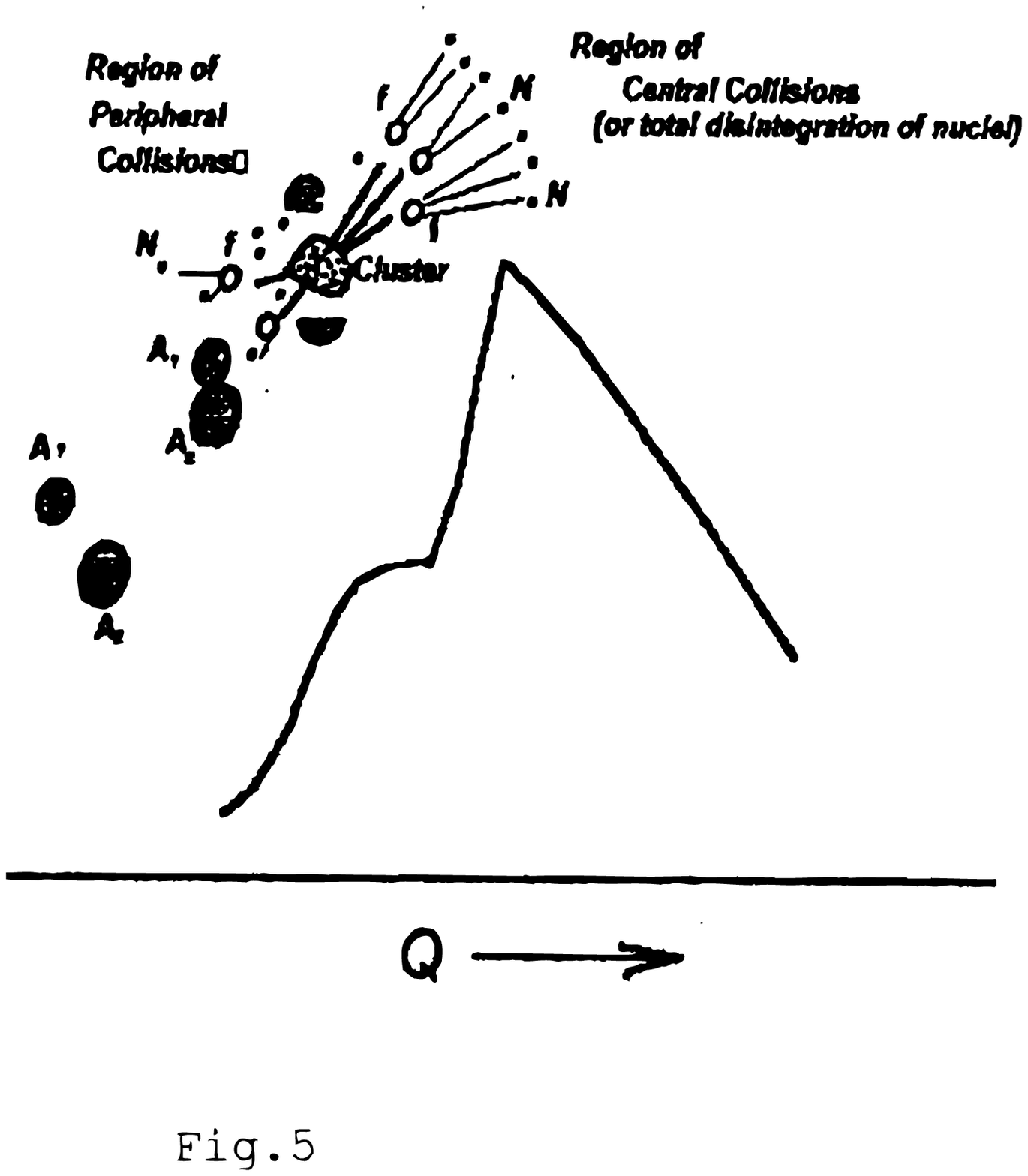}{}}

\vskip 0.25cm

\end{center}

\end{document}